\documentclass[useAMS,usenatbib]{mn2e}
\usepackage{times}

\title[High resolution imaging of CLASS B2045+265]{High resolution imaging of the anomalous flux-ratio gravitational lens system CLASS B2045+265: Dark or luminous satellites?}
\author[J. P. McKean et al.]{J. P. McKean,$^{1,2}$\thanks{Email: mckean@mpifr-bonn.mpg.de} L. V. E. Koopmans,$^3$ C. E. Flack,$^{1,4}$ C. D. Fassnacht,$^1$ D. Thompson,$^5$
\newauthor K. Matthews,$^5$ R. D. Blandford,$^6$ A. C. S. Readhead$^7$ and B. T. Soifer$^5$\\
$^1$Department of Physics, University of California, Davis, CA 95616, USA\\
$^2$Max-Planck-Institut f\"{u}r Radioastronomie, Auf dem H\"{u}gel 69, D-53121 Bonn, Germany\\
$^3$Kapteyn Astronomical Institute, Postbus 800, NL-9700 AV Groningen, the Netherlands\\
$^4$Bemidji State University, Bemidji, MN 56601, USA\\
$^5$Caltech Optical Observatories, California Institute of Technology, Pasadena, CA 91125, USA\\
$^6$KIPAC, Stanford University, 2575 Sand Hill Road, Menlo Park, CA 94025, USA\\
$^7$Department of Astronomy, California Institute of Technology, Pasadena, CA 91125, USA\\}
\begin{document}

\date{Accepted 2007 March 16. Received 2007 March 15; in original form 2006 October 23}

\pagerange{\pageref{firstpage}--\pageref{lastpage}} \pubyear{2005}

\maketitle

\label{firstpage}

\begin{abstract}
The existence of flux-ratio anomalies between fold and cusp images in galaxy-scale strong-lens systems has led to an interpretation based on the presence of a high mass-fraction of cold-dark-matter (CDM) substructures around galaxies, as predicted by numerical N-body simulations.  These substructures can cause large perturbations of the image magnifications, leading to changes in the image flux ratios. The flux-ratio anomaly is particularly evident in the radio-loud quadruple gravitational lens system CLASS B2045+265.  In this paper, new high-resolution radio, optical, and infrared imaging of B2045+265 is presented which sheds more light on this anomaly and its possible causes. First, deep Very Long Baseline Array observations show very compact images, possibly with a hint of a jet, but with no evidence for differential scattering or scatter broadening. Hence, the flux-ratio anomaly is unlikely to be caused by refractive scattering in either the Milky Way or the lens galaxy. Second, optical and infrared observations with the {\it Hubble Space Telescope} and through Adaptive-Optics imaging with the W. M. Keck Telescope, show a previously undiscovered object -- interpreted as a (tidally disrupted) dwarf satellite based on its colours and slight extension -- between the main lens galaxy and the three anomalous flux-ratio images. Third, colour variations in the early-type lens galaxy indicate recent star-formation, possibly the result of secondary infall of gas-rich satellites. A population of such galaxies around the lens system could explain the previously discovered strong [O {\sc ii}] emission. However, spiral structure and$/$or normal star formation in the lens galaxy cannot be excluded. In light of these new data, we propose a lens model for the system, including the observed dwarf satellite, which reproduces all positional {\it and} flux-ratio constraints, without the need for additional CDM substructure. Although the model is peculiar in that the dwarf galaxy must be highly flattened, the model is very similar to recently proposed mass models based on high-order multipole expansions.
\end{abstract}


\section{Introduction}

The standard model of cosmological structure formation predicts that galaxies form hierarchically from the mergers of smaller-mass progenitors, leading to increasingly more massive galaxies.  According to this model -- as seen in numerical simulations -- the baryonic stellar and gas components of galaxies should be surrounded by an extended ``smooth'' dark-matter component, containing many dwarf satellites. In cold-dark-matter (CDM) only simulations, this ensemble of compact satellites is generically termed ``CDM substructure'' (\citealt*{navarro96}; \citealt{moore98}). However, if one compares the density of these satellites in numerical simulations with those found around, e.g., the Milky Way or M31, a severe discrepancy is apparent in the number of observed dwarf satellites. There are 1--2 orders of magnitude fewer dwarfs observed than predicted \citep{klypin99,moore99}. If this discrepancy is genuine, then either these satellites do not exist and the $\Lambda$CDM paradigm has a major problem, or they consist purely of dark matter having had their baryons stripped during the early phases of the galaxy-formation process.

\begin{table*}
\begin{center}
\caption{The flux-densities, positions and angular sizes of the B2045+265 radio components, as measured with the VLBA at 5~GHz. The errors in the flux densities include a 5 per cent uncertainty in the absolute flux-density calibration of the VLBA. The flux density ratios are measured relative to image A (A1$+$A2). The position of component A1 is 20$^h$47$^m$20.28593$^s$ $+$26$\degr$44$\arcmin$02.6911$\arcsec$.}
\begin{tabular}{lrrrrcccc} \hline
Comp.	& \multicolumn{1}{c}{$\Delta\alpha$} & \multicolumn{1}{c}{$\Delta\delta$} & \multicolumn{1}{c}{$I_{peak}$}        & \multicolumn{1}{c}{$S_{int}$} & Major Axis   & $\sigma_{map}$ & Flux density ratio\\
	& \multicolumn{1}{c}{[mas]}	 & \multicolumn{1}{c}{[mas]}	  & \multicolumn{1}{c}{[mJy~beam$^{-1}$]} & \multicolumn{1}{c}{[mJy]}     & [mas] & [$\mu$Jy~beam$^{-1}$] & [$S_{int}/S_A$] \\ \hline
A1	&       0.0$\pm$0.1   &       0.0$\pm$0.1      & 12.68$\pm$0.64  & 13.89$\pm$0.70  & 1.1   & 58  & 0.95$\pm$0.01\\
A2	&    $-$1.6$\pm$0.5   &    $-$2.2$\pm$0.5      &  0.67$\pm$0.07  &  1.19$\pm$0.18  & 4.1   & 58  & 0.05$\pm$0.01\\
B	&  $-$133.8$\pm$0.1   &  $-$248.3$\pm$0.1      &  8.48$\pm$0.43  &  9.17$\pm$0.47  & 1.0   & 64  & 0.61$\pm$0.01\\
C	&  $-$287.7$\pm$0.1   &  $-$790.4$\pm$0.1      & 12.99$\pm$0.65  & 14.04$\pm$0.71  & 0.9   & 54  & 0.93$\pm$0.01\\
D	& $+$1626.8$\pm$0.2   & $-$1006.4$\pm$0.2      &  1.47$\pm$0.09  &  1.60$\pm$0.13  & 0.9   & 56  & 0.11$\pm$0.01\\
E	& $+$1107.4$\pm$0.2   &  $-$807.1$\pm$0.2      &  1.72$\pm$0.10  &  1.76$\pm$0.13  & 0.7   & 51  & 0.12$\pm$0.01\\
\hline
\end{tabular}
\end{center}
\label{rad-tab}
\end{table*}

Strong gravitational lensing offers a powerful method to detect and quantify the presence of mass substructure, be it luminous or dark. This detection could be achieved through the detailed modelling of Einstein ring emission (e.g. \citealt{koopmans05}).  Alternatively, substructure may also be detected through its effect on the flux-ratios between fold or cusp lensed images (\citealt{mao98}; \citealt*{keeton03,keeton05}; \citealt{bradac04}). When the source distance to the fold or cusp is much smaller than the Einstein radius, the sum of the image brightnesses (assuming they are point images and not affected by variability) add to zero when the sign of the image parities are accounted for (i.e.\ negative-parity images are taken to have negative brightness and magnification). Most lens systems with fold and cusp images appear not to obey these relationships between image brightnesses, which in principle are independent of the global mass model. These flux-ratio anomalies have subsequently been interpreted as the first direct evidence for the existence of CDM substructure around lensing galaxies (\citealt{bradac02,chiba02,dalal02,metcalf02,keeton03,kochanek04,bradac04,keeton05}). However, the mass fraction of this substructure appears to be on the high side when compared to models \citep{mao04} and the precise interpretation of these intriguing results remains open for debate.

In this paper, we present new optical, near-infrared and radio data on the gravitational lens system CLASS B2045+265, which has one of the largest flux-ratio anomalies of all known lens systems. These data, and in particular the discovery of a previously unknown optically-bright dwarf-galaxy between the main lensing galaxy and the three cusp images, sheds new light on whether the anomaly is caused by differential scattering, genuine dark-matter-only satellites or possibly by luminous dwarf satellites. In Section \ref{review}, we give a brief description of the previous observations and models for B2045+265, and discuss the possible reasons for the flux ratio anomaly. Our new high-resolution radio data are presented in Section \ref{radio}. The optical and infrared data sets, which include new adaptive optics imaging with the W. M. Keck Telescope and archival {\it Hubble Space Telescope} data, are presented in Section \ref{optical}. The radio, optical, and infrared data are combined and used to constrain a new mass model for B2045+265, which includes the presence of the newly-discovered dwarf galaxy, in Section \ref{model}. Finally, in Section \ref{conclusions} we discuss our results and present our conclusions. Throughout this paper, we adopt an $\Omega_{M} =$~0.3, $\Omega_{\Lambda} =$~0.7 flat Universe, with a Hubble parameter of $H_{0} =$~100~$h$~km~s$^{-1}$~Mpc$^{-1}$.

\begin{figure*}
\begin{center}
\setlength{\unitlength}{1cm}
\begin{picture}(12,9.5)
\put(-1.0,-0.05){\includegraphics{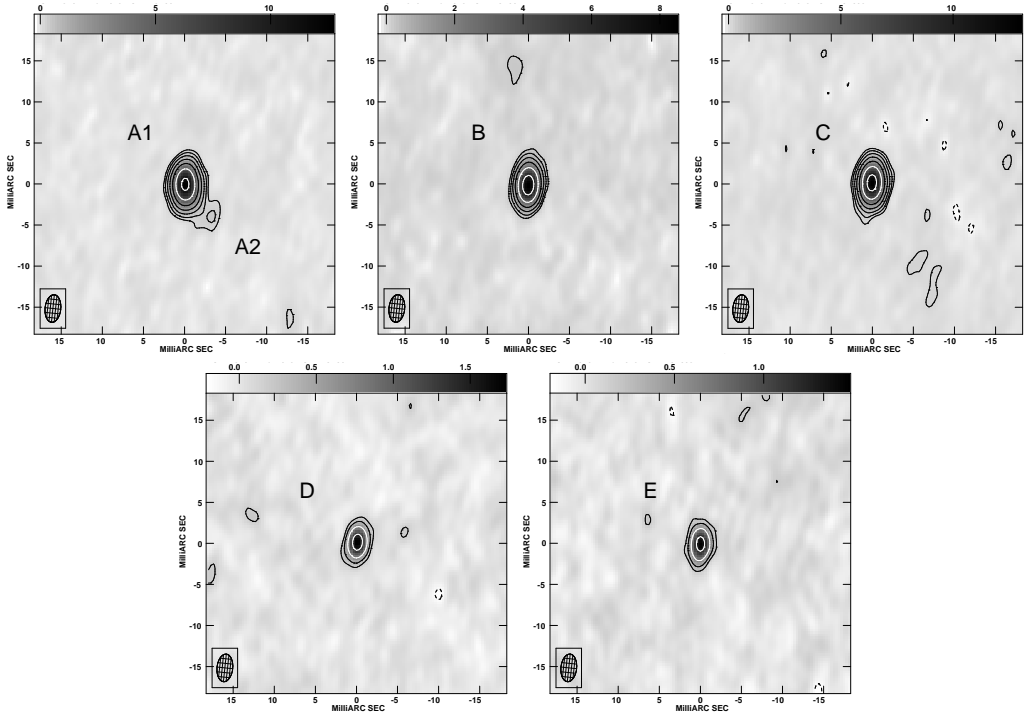}}
\end{picture}
\caption{The naturally weighted 5-GHz maps of the B2045+265 lensed images (A, B, C and D) and lensing galaxy (E) taken with the VLBA on 1997 October 17. Lensed image A is resolved into a compact core (A1) with a faint extended jet to the south-west (A2). The remaining lensed images are compact, although there is evidence for a slight extension in image C. The weak and compact radio emission from component E is consistent with an AGN embedded within the lensing galaxy, G1. The contours in each map are ($-$3, 3, 6, 12, 24, 48, $\ldots$)~$\times$~$\sigma_{map}$, the rms map noise given in Table~\ref{rad-tab}. The restoring beam size is 3.4$\times$1.9~mas$^2$ at a PA of $-$5.5$\degr$. The grey-scales are in units of mJy~beam$^{-1}$. North is up and east is left.}
\label{vlba-lensed}
\end{center}
\end{figure*}

\section{CLASS B2045+265}
\label{review}

\citet{fassnacht99a} report the discovery of B2045+265, which is one of the most intriguing gravitational lens systems to be found during the course of the Cosmic Lens All-Sky Survey (CLASS; \citealt{myers02}; \citealt{browne02}). This system comprises four radio-loud lensed images in a long-axis cusp configuration; the three brightest lensed images (A, B and C) being almost co-linear in a north--south direction, and very close together on the sky (0.84~arcsec separation). The fourth and faintest lensed image, D, is 1.91~arcsec from the brightest lensed image, A, and is $\sim$9 times fainter at 5~GHz. The system also has emission from a fifth radio component, E, which lies between D and the other lensed images. Multi-frequency radio mapping from 1.4 to 15 GHz with the Very Large Array (VLA) and Multi-Element Radio Linked Interferometric Network (MERLIN) found the radio spectra of the four lensed images to be nearly identical ($\alpha_{1.4}^{15}\sim$~$-$0.42 where $S_{\nu} \propto \nu^{\alpha}$). However, the radio spectrum and flux-density of component E was not consistent with a fifth lensed image. Therefore, E is believed to be emission from an AGN (Active Galactic Nucleus) within the lensing galaxy. Follow-up VLBA (Very Long Baseline Array) imaging at 5~GHz detected unresolved emission from lensed images A, B and C. However, the short 35-min observation was not sensitive enough to detect image D or radio component E.

High resolution infrared imaging of B2045+265 with the Near Infrared Camera/Multi-Object Spectrograph (NICMOS) on board the {\it Hubble Space Telescope} found a lensing galaxy coincident with the emission from radio component E and strong point source emission from the lensed images A, B and C with an underlying gravitational arc. The weakest lensed image was not detected in the infrared. Follow-up spectroscopic observations of the lensing galaxy showed strong [O~{\sc ii}] emission and Balmer absorption lines which together are consistent with an Sa late-type galaxy at $z =$~0.867 \citep{fassnacht99a}. A single, broad and isolated emission line was found in the spectrum of the background quasar, which has been identified as Mg {\sc ii} at a redshift of $z=$~1.28. However, the uncomfortably low source redshift with respect to the lens requires a massive deflector, with a large dark matter component, to cause the 1.91~arcsec image splitting observed in B2045+265 (see \citealt{fassnacht99a}). Recently, \citet{hamana05} measured the stellar velocity dispersion of the lensing galaxy to be $\sigma_{v}=$~213$\pm$23~km~s$^{-1}$. This is significantly higher than the characteristic velocity dispersion for the late-type lensing galaxy population obtained from the CLASS gravitational lensing statistics ($\sigma_v=$~117$^{+45}_{-31}$; \citealt{chae02}), leading to some doubt over the Sa-type classification of the lensing galaxy. Furthermore, the joint lensing and dynamics analysis carried out by \citeauthor{hamana05} concluded that the overall density profile, $\rho(r)$, of the lens must be shallower than isothermal ($\beta =$~1.58 where $\rho(r) \propto r^{-\beta}$), which may be due in part to a possible nearby group of galaxies not accounted for in the modelling \citep{fassnacht99a}. Conversely, if the source redshift is higher than the 1.28 reported by \citeauthor{fassnacht99a}, then the lens could have a lower mass and hence a steeper density profile, consistent with the measured stellar velocity dispersion.

Attempts to model the system with a smooth mass distribution have successfully reproduced the positions of the lensed images to within the astrometric uncertainties, but failed to account for the flux ratios (e.g. \citealt{fassnacht99a}). For a smooth global mass model, B is expected to be the brightest of the three merging lensed images; however, image A is significantly brighter than image B. The flux ratio anomaly of B2045+265 can be best quantified using the dimensionless $R_{cusp}$ parameter \citep{bradac02,keeton03},
\begin{equation}
R_{cusp} = \frac{\mu_A + \mu_B + \mu_C}{|\mu_A| + |\mu_B| + |\mu_C|},
\end{equation}
where $\mu_A$, $\mu_B$ and $\mu_C$ are the magnifications of merging images A, B and C, respectively. For a true cusp lens system with a globally smooth mass distribution, $R_{cusp}$ should tend to 0 as the source position approaches the cusp, i.e., the flux density of B should approach the sum of the flux densities of A and C. In the case of B2045+265, $R_{cusp}=~$0.506$\pm$0.013 \citep{koopmans03}, which is one of the most extreme examples of a cusp flux ratio anomaly known \citep{keeton03,metcalf05}. Furthermore, the large value for $R_{cusp}$ is mostly due to images B and C both being much fainter (by 25--50~per cent) than what is predicted from a simple mass model \citep{dobler06}. This could be accounted for if there were several CDM substucture clumps within the lensing galaxy or along the line-of-sight to images B and C.  However, this interpretation may be inconsistent with the level of substructure predicted from $\Lambda$CDM simulations.

Alternatively, the flux ratio anomaly may be due to propagation effects, such as differential scattering or free-free absorption. Both of these effects are believed to be responsible for changing the properties of radio-loud lensed images in other gravitational lens systems (e.g. \citealt*{winn03}; \citealt{biggs04}; \citealt*{mittal07}). However, free-free absorption and refractive scattering are frequency dependent and the radio spectra of the four lensed images of B2045+265 are nearly identical from 1.4 to 15~GHz \citep{fassnacht99a}. Therefore, the cause of the flux ratio anomaly in B2045+265 is almost certainly gravitational in origin.

\section{Radio Observations}
\label{radio}

In this section, we present new high resolution radio observations of B2045+265, obtained with the VLBA.

\subsection{VLBA 5-GHz observation}

The snapshot VLBA imaging presented by \citet{fassnacht99a} was only sensitive enough to detect the three brightest lensed images (A, B and C) of B2045+265 at 5 GHz. Therefore, with the aims of detecting the weakest lensed image (D) at mas-scales, investigating the nature of the fifth radio component (E), and searching for extended structure in all of the lensed images, we undertook a deep observation of B2045+265 with the VLBA at 5 GHz. The observation was phase referenced and carried out over 10.7~h on 1997 October 17, with $\sim$6.5~h spent on B2045+265. The data were recorded in the left circular polarisation through 4 IFs, each with a bandwidth of 8-MHz. A bit rate of 128~Mb~s$^{-1}$ with 2-bit sampling was employed. Phase referencing and fringe finding were carried out with JVAS B2048+312 \citep{wilkinson98} and 3C 454.3, respectively. The data were correlated at Socorro, where the individual IFs were divided into 16$\times$0.5~MHz channels and the visibilities were averaged over 1-s time intervals.

The data editing and calibration was performed within {\sc aips} (Astronomical Image Processing Software) using the {\sc vlbapipe} pipeline script. No further frequency or time averaging was undertaken during this process. This produced an effective field of view for the resulting maps of $\sim$5 arcsec from the phase tracking centre, easily large enough to contain the full extent of the lens system, which has a maximum lensed image separation of 1.91 arcsec. The image deconvolution was carried out using {\it imagr}. We first created a shallow wide-field map and identified the locations of the five radio components of B2045+265. A small 80$\times$80~mas$^{2}$ field was then placed around each component. A process of {\it clean}ing with two cycles of phase-only self calibration, using a solution interval of 10~min, was carried out. Natural weighting was used throughout. The resulting restoring beam size was 3.4$\times$1.9~mas$^{2}$ at a PA of $-$5.5$\degr$. The image rms in each field was $\sim$55~$\mu$Jy~beam$^{-1}$. Finally, elliptical Gaussian model components were fitted to the radio components using {\it jmfit}.

\subsection{Results}

The deconvolved naturally-weighted maps of the four lensed images are presented in Fig. \ref{vlba-lensed} and show a mixture of compact and extended emission consistent with gravitational lensing. Image A has been resolved into two subcomponents, forming a core-jet structure. Subcomponent A1, the core, is compact and is the dominant source of the 5 GHz emission from image A. The faint extended nature of subcomponent A2 is consistent with a radio jet. Furthermore, A2 is extended toward the south west, which is perpendicular to the axis connecting image A to the lensing galaxy. In contrast, image B is unresolved in our deep VLBA map. This is not surprising since the surface brightness of the lensed images is conserved by gravitational lensing and image B is fainter than image A (the $S_B/S_A$ flux density ratio is 0.61). The emission from image C is well represented by a single compact elliptical Gaussian. However, it is evident from Fig. \ref{vlba-lensed} that there is a slight extension towards the south in image C, which is presumably due to faint extended emission from the lensed jet. We have successfully detected the fourth lensed image of B2045+265 for the first time at mas-scales. Image D is the weakest lensed image and has a compact structure. The results of fitting Gaussian model components to the B2045+265 lensed images are presented in Table \ref{rad-tab}.

We have also detected the fifth radio component of B2045+265 with the VLBA. The naturally weighted map is also shown in Fig. \ref{vlba-lensed}. We find the component E radio emission to be both weak and unresolved. \citet{fassnacht99a} argue that component E is most probably emission from the lensing galaxy because the flux density is too high to be a core lensed image.\footnote{Recall that component E also has a flatter radio spectrum than the lensed images.} We again find this to be the case, with component E having a higher flux density than image D. Assuming that component E is associated with the lensing galaxy, we calculate the rest-frame 1.4~GHz luminosity of the component E radio emission to be 2.9~$\times$~10$^{23}$~$h^{-2}$~W~Hz$^{-1}$, which is consistent with a weak AGN. The flat radio spectrum ($\alpha_{5}^{15} \sim -0.4$ between 5 and 15~GHz, where $S_{\nu} \propto \nu^{\alpha}$) and compact nature of component E at mas-scales are both in agreement with the classification as a radio-loud AGN. The properties of the elliptical Gaussian fitted to component E are also presented in Table~\ref{rad-tab}.

\section{Optical and Infrared Observations}
\label{optical}

In this section, we present new adaptive optics imaging acquired with the W. M. Keck Telescope, and archival data taken with the {\it Hubble Space Telescope}.

\begin{table*}
\begin{center}
\caption{The relative positions and (Vega) magnitudes of the optical and infrared components of B2045+265. Note that radio component E is coincident with G1. The relative positions have been taken from the 2.2~$\mu$m adaptive optics imaging. The magnitudes have not been corrected for galactic extinction. Non-dectections are quoted at the $3~\sigma$ level.}
\begin{tabular}{lrrccccc} \hline
Comp.	& \multicolumn{1}{c}{$\Delta\alpha$} & \multicolumn{1}{c}{$\Delta\delta$} & F555W & F814W & F160W & {\it K}\\
	& \multicolumn{1}{c}{[mas]}	     & \multicolumn{1}{c}{[mas]}	  & \multicolumn{4}{c}{[Vega magnitudes]} \\ \hline
A	&       0.0$\pm$0.5    &       0.0$\pm$0.5      & 25.56$\pm$0.12 & 22.97$\pm$0.05 & 19.94$\pm$0.04 & 18.29$\pm$0.17 \\
B	&  $-$131.6$\pm$0.6    &  $-$244.8$\pm$0.6      & 26.10$\pm$0.17 & 23.61$\pm$0.06 & 20.40$\pm$0.04 & 18.74$\pm$0.17\\
C	&  $-$286.9$\pm$0.5    &  $-$788.5$\pm$0.5      & 25.72$\pm$0.12 & 23.21$\pm$0.06 & 20.07$\pm$0.04 & 18.52$\pm$0.15\\
D	& $+$1626.8$\pm$1.3    & $-$1006.4$\pm$1.3      & $>$27.14	 & $>$25.77       & 22.52$\pm$0.05 & 20.57$\pm$0.15\\
G1	& $+$1108.4$\pm$1.1    &  $-$806.5$\pm$1.1      & 23.54$\pm$0.07 & 21.02$\pm$0.04 & 18.28$\pm$0.07 & 17.38$\pm$0.15\\
G2	&  $+$449.8$\pm$2.1    &  $-$642.5$\pm$2.1      & $>$27.14	 & 24.65$\pm$0.08 & 22.84$\pm$0.08 & 21.90$\pm$0.16\\
\hline
\end{tabular}
\label{opt-tab}
\end{center}
\end{table*}

\begin{figure}
\begin{center}
\setlength{\unitlength}{1cm}
\begin{picture}(6,5.40)
\put(-1.2,-0.15){\includegraphics{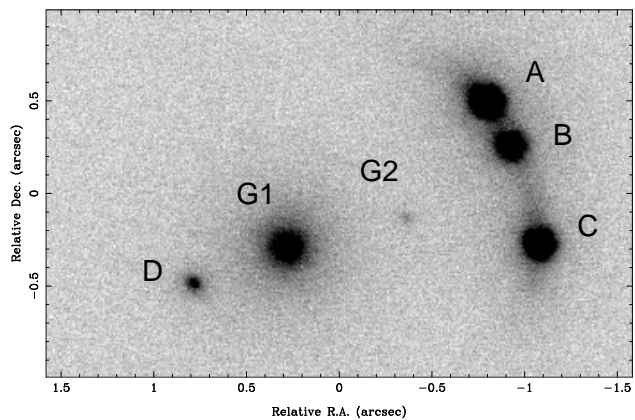}}
\end{picture}
\caption{Adaptive optics imaging at 2.2~$\mu$m with NIRC2 on the W. M. Keck-II Telescope. These new high resolution data have detected for the first time a faint galaxy (G2) which may be responsible for the flux ratio anomaly observed in B2045+265. North is up and east is left.}
\label{ao}
\end{center}
\end{figure}

\subsection{W. M. Keck Telescope adaptive-optics imaging}
\label{ao-section}

High resolution ground based imaging of B2045+265 at 2.2~$\mu$m ({\it K}-band) was obtained on 2005 July 31 with the Near InfraRed Camera 2 (NIRC2; Matthews et al. in preparation) behind the adaptive-optics bench on the W. M. Keck-II Telescope. The NIRC2 narrow camera was used throughout, which provides a field of view of 10~$\times$~10 arcsec$^2$ and a pixel-scale of 9.94~mas~pixel$^{-1}$. The data were taken in twenty three 180-s exposures with a small dither between each to facilitate good sky background subtraction during the reduction stage. A sodium laser guide star was used to correct for the atmospheric turbulence, and a $V=$~14.1-mag natural guide star, 33~arcsec from B2045+265, was used for fast guiding.

The data were reduced within {\sc iraf}\footnote{{\sc iraf} (Image Reduction and Analysis Facility) is distributed by the National Optical Astronomy Observatories, which are operated by AURA, Inc., under cooperative agreement with the National Science Foundation.} using a double-pass reduction algorithm. The first pass is used to map the positions of the objects on the sky, which are then masked out during the second pass where the usual sky subtraction with temporally adjacent frames is carried out. The bad pixel and cosmic ray masks generated during this process were then used to construct weight maps. The individual exposures were combined using the weight maps and the {\it drizzle} package \citep{fruchter02}, which also corrected the data for the geometric distortions across the NIRC2 camera. We found that the quality of the corrected images varied greatly over the dataset we obtained on B2045+265, with the point-source full width at half maximum (FWHM) ranging from 67 to 93~mas. Therefore, we also drizzled subsets of the data together, which included the 6 and 12 best individual exposures i.e. those with the sharpest point spread function (psf).

In Fig. \ref{ao} we show our adaptive optics assisted {\it K}-band image constructed using all twenty-three exposures of B2045+265 (psf FWHM of $\sim$93~mas). First, it is clear that all of the radio components found in the VLBA maps have infrared counterparts (note that radio component E has been labelled G1 in Fig. \ref{ao}). Strong point-source emission from the four lensed images has been detected. Also, the infrared gravitational arc associated with A, B and C has once again been observed. The lensing galaxy G1, which is coincident with radio component E to within the positional uncertainties, has a circular morphology (axial ratio of $b/a =$~0.94$\pm$0.01 at a position angle of $-$60.4$\degr$$\pm$4.8$\degr$ east of north) with no evidence of spiral arms. Furthermore, the surface brightness profile of G1, measured using {\sc galfit} \citep{peng02}, is consistent with a de Vaucouleurs model with an effective radius of 0.465~arcsec ($\equiv$~1.65~$h^{-1}$~kpc). This result is not too surprising since these {\it K}-band observations correspond to 1.2~$\mu$m in the rest frame of G1, and are therefore sensitive to the older (red) stellar population contained within the bulge. Finally, and most importantly, we have also detected infrared emission from a previously unknown object (which we label G2), positioned between the main lensing galaxy G1 and the lensed images A, B and C. G2 is significantly fainter than G1 (the G1$/$G2 flux ratio is $\sim$~64). In Section \ref{model} we will test whether G2 is responsible for the large flux ratio anomaly between the three merging images of B2045+265.

In Table \ref{opt-tab} we report the relative positions and apparent magnitudes for the infrared components of B2045+265. No photometric standard stars were observed during our adaptive optics observations. Therefore, we compared the brightnesses of stars in our adaptive optics imaging and the calibrated NIRC {\it K}-band imaging presented by \citet{fassnacht99a} to calculate the NIRC2 zeropoint magnitude -- note that there were only two isolated stars common to both images. The absolute photometric accuracy resulting from this process is estimated to be no better than $\sim$~0.15 mags. Also, the relative photometric accuracy from our adaptive optics imaging will be limited by changes in the psf, and therefore the Strehl ratio, over time and position on the detector (e.g. \citealt{esslinger98}). For G1 we state the total magnitude from the fitted de Vaucouleurs profile. A circular aperture\footnote{The radii of all circular apertures used throughout this paper are set such that, for a point source, the measured magnitudes from aperture photometry and psf fitting would be identical.} was used to measure the magnitudes of the lensed images and G2, from a G1 subtracted image.

\begin{figure*}
\begin{center}
\setlength{\unitlength}{1cm}
\begin{picture}(12,5.75)
\put(-2.70,-0.15){\includegraphics{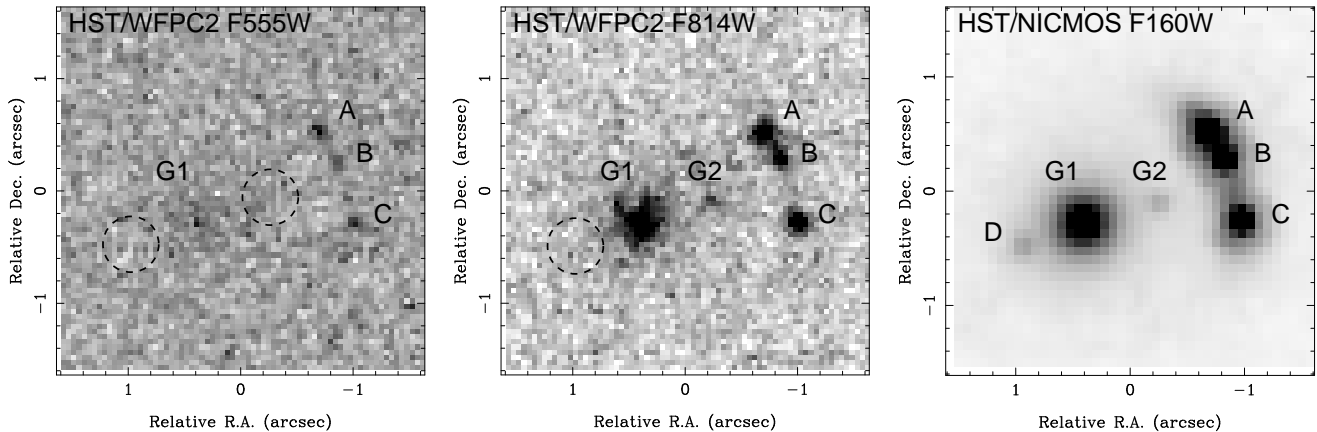}}
\end{picture}
\caption{Optical and infrared imaging of B2045+265 with WFPC2 and NICMOS on the {\it Hubble Space Telescope}. The WFPC2 F555W and F814W data were not sensitive enough to detect all the components of B2045+265. A dashed circled has been placed over the positions of the missing components. North is up and east is left in each image.}
\label{hst-image}
\end{center}
\end{figure*}

\subsection{Hubble Space Telescope imaging}
\label{hst}

To investigate the nature of G2 further, we have retrieved high resolution optical and infrared imaging of B2045+265 from the {\it Hubble Space Telescope} archive. The optical data were taken with the Wide Field Planetary Camera (WFPC2) on 1998 December 04 and consist of two 500-s exposures through both the F555W and F814W filters (GO-6629; PI: Jackson). The lens system was positioned in the centre of the Planetary Camera (PC1), which has a field of view of 37~$\times$~37~arcsec$^{2}$ and a pixel scale of 46~mas~pixel$^{-1}$. B2045+265 was re-observed with NICMOS on 2003 September 6 (G0-9744; PI: Kochanek). These new observations were much deeper than those presented by \citet{fassnacht99a}, and were designed to detect the fourth lensed image and provide high signal to noise ratio imaging of the lensing galaxy. The data were acquired with the F160W filter and the NIC2 camera, which has a field of view of 19~$\times$~19~arcsec$^2$ and a pixel scale of 75~mas~pixel$^{-1}$. The imaging was taken in five 640-s and three 704-s exposures (total time 5.3~ks).

The WFPC2 and NICMOS datasets were downloaded from the Multimission Archive at Space Telescope and processed within {\sc iraf} using standard {\sc stsdas} procedures. In particular, the {\it multidrizzle} package \citep{koekemoer02} was used to combine the individual exposures in each filter, remove cosmic ray events and correct for the WFPC2 and NICMOS geometric distortions. Our final optical and infrared images of B2045+265 are presented in Fig. \ref{hst-image}. Note that the WFPC2 imaging was not sensitive enough to detect all of the components of B2045+265. Image D was not detected at either F555W or F814W. Furthermore, G2 was not found in the F555W image and the main lensing galaxy G1 was only marginally detected. These non-detections are due in part to the short exposure times through both the F555W and F814W filters with WFPC2, but are also due to the high extinction along the line-of-sight to B2045+265 owing to its low galactic latitude ($b=-$10.4$\degr$; $A_V=$~0.8~mags; \citealt{schlegel98}). The positions of the missing components in the F555W and F814W images have been marked by the dashed circles in Fig.\ref{hst-image}.

The structure of the lensing galaxy is significantly different in the F814W and F160W images. At F814W, there is clear evidence of an irregular morphology which would be consistent with emission from either spiral arms or additional infalling dwarf galaxies. Alternatively, the observed structure could be artifacts of the data reduction process (we cannot rule out the possibility that these features are residual cosmic rays or bad pixels as only two undithered observations of B2045+265 were taken with WFPC2). Conversely, the lensing galaxy has a smooth surface brightness distribution at F160W, which is well fitted with a de Vaucouleurs profile. The fitted effective radius (0.411~arcsec) and axial ratio (0.92$\pm$0.01 at a position angle of $-$66.0$\degr$$\pm$4.3$\degr$ east of north) of G1 at F160W are both consistent with the results found using the {\it K}-band adaptive optics imaging presented in Section \ref{ao-section}. The lensed images of the background quasar are unresolved point sources at F555W, F814W and F160W (note that the gravitational arc associated with A, B and C was found at F160W after subtracting the lensed images -- but is not shown here). The flux ratio anomaly between the three merging lensed images is also present in the optical and infrared (e.g. at F160W the flux-ratios are $F_B/F_A=$~0.65$\pm$0.04 and $F_C/F_A=$~0.89$\pm$0.04), which further adds to the case for the amomaly being gravitational in origin. In an attempt to characterise the surface brightness properties of G2, we have fitted the F160W data for G2 with a simple Gaussian profile. We find the G2 emission to be slightly extended and elongated toward the lensed images. The fitted Gaussian for G2 has an axial ratio of 0.56$\pm$0.18 at a position angle of 106.3$\degr$$\pm$21.1$\degr$ east of north. The G1$/$G2 flux ratio at F160W is $\sim$~67.

In Table \ref{opt-tab} we give the optical and infrared magnitudes of G1, G2 and the lensed images. For G1, we report the total magnitude for the best fitting de Vaucouleurs profile (for the F555W and F814W datasets, we have held fixed the parameters of the de Vaucouleurs profile fitted to the F160W data, and allowed only the total magnitude to vary). Point-sources were used to determine the magnitudes of the lensed images through each filter. For G2, we have measured the magnitude within a circular aperture from an image with the lensing galaxy and lensed images subtracted.

\subsection{The nature of G2}

We have labelled the newly detected object G2 which implies that it is a second galaxy along the line-of-sight to B2045+265. However, given the low galactic latitude of B2045+265 there is the possibility that G2 may be a star. The strongest evidence for G2 being a galaxy comes from the extension and elongation detected in the F160W imaging. We have compared the size of G2 with the size of 10 stellar objects in the F160W field (the F160W imaging has the highest signal-to-noise ratio for G2 of all the available datasets); the average deconvolved FWHM of the stars is 0.73 pixels (rms of 0.22), whereas for G2 the deconvolved FWHM is 1.71$\pm$0.30 pixels.

Measuring the spectrum and obtaining a redshift for G2 would provide conclusive evidence that it is indeed a galaxy. Extensive spectroscopic observations of the system by \citet{fassnacht99a} were taken at a position angle (112$\degr$) which also placed the slit over G2. The resulting spectrum showed strong [O~{\sc ii}] emission at the same redshift as lensing galaxy G1, which could be due to star-formation associated with either G1 or G2.  However, these ground based spectroscopic observations did not spatially resolve G2 from G1, and therefore the exact source of the [O~{\sc ii}] emission is unknown.

\begin{figure}
\begin{center}
\setlength{\unitlength}{1cm}
\begin{picture}(6,6)
\put(-1.9,6.6){\includegraphics{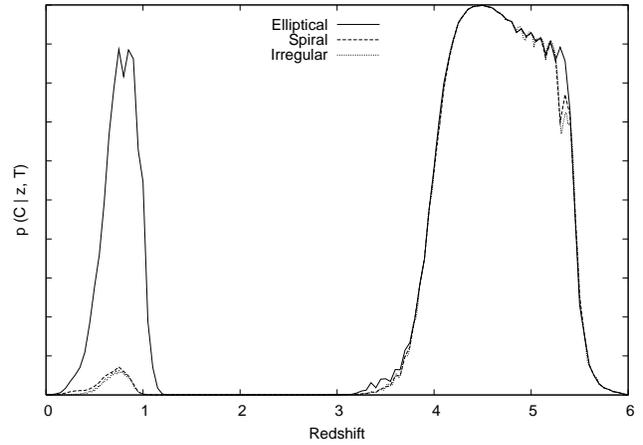}}
\end{picture}
\caption{The G2 photometric redshift probability distribution for elliptical, spiral and irregular galaxy templates found using {\sc hyperz} and the optical and infrared data presented here. Due to the limited number of photometric data points available, there is a two peaked probability--redshift distribution. The photometric redshift of the two peaks are $z_{phot} \sim$~0.86 and 4.50.}
\label{g2-pz}
\end{center}
\end{figure}

In an attempt to constrain the redshift of G2, we have calculated a photometric redshift using the spectral energy distribution fitting code {\sc hyperz} \citep*{bolzonella00}. This code compares the observed spectral energy distribution with an ensemble of theoretical$/$observed template spectra to establish the most probable redshift and spectral type. The optical and infrared data (C) presented in Table \ref{opt-tab}, after correcting for galactic extinction using the $E_{B-V}$ reddening maps of \citet{schlegel98} and the $R_{V}$-dependent extinction law of \citet*{cardelli89}, provided the observed photometric properties of G2. The template spectra (T) for elliptical (instantaneous burst, E$/$S0), spiral (Sa--Sd) and irregular (Ir) galaxy types with 51 different stellar population ages were taken from the \citet{bruzual93} GISSEL98 spectral library. The resulting G2 probability--redshift distribution for each template, p(C$|z$,T), is given in Fig. \ref{g2-pz}. 

\begin{figure}
\begin{center}
\setlength{\unitlength}{1cm}
\begin{picture}(6,6)
\put(-1.9,6.6){\includegraphics{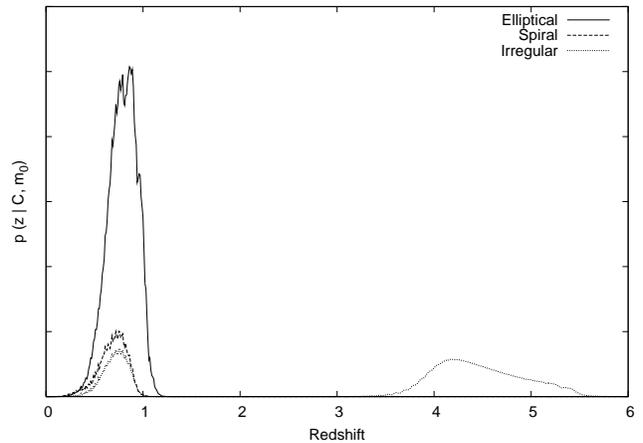}}
\end{picture}
\caption{The photometric redshift probability distribution for G2, with the addition of a Bayesian prior on the apparent magnitude. The best fitting template for G2 is an elliptical at $z_{phot}=$~0.86$^{+0.08}_{-0.21}$.}
\label{g2-pz-bay}
\end{center}
\end{figure}

There are two clear peaks in the probability--redshift distribution for each galaxy template; the first is at $z_{phot}\sim$~0.86 and the second, which is much broader in redshift space, is at $z_{phot}\sim$~4.50. This double-peaked probability--redshift distribution is a typical result from photometric redshift codes when there is insufficient coverage of the observed spectral energy distribution. This is because strong spectral features such as the 4000~{\AA} and Lyman breaks are used to determine the photometric redshift and it is not always clear which break has been detected from only a few photometric measurements. To overcome this, we have also applied a Bayesian prior using the apparent magnitude ($m_{0}$) of G2. Here, we follow the procedure described by \citet{benitez00} where redshift distributions per spectral type are calculated from Hubble Deep Field-North data and used to establish the relative redshift--probability for a given template and apparent magnitude. This process essentially places a luminosity limit per spectral type e.g. there are fewer $L^*$ ellipticals at redshift 5 than at redshift 1, thus downweighting any high redshift peak. In Fig. \ref{g2-pz-bay} we show the probability-redshift distribution for G2 using the prior on the apparent magnitude, p($z|$C,$m_{0}$). We see that for the elliptical and spiral spectral types the high redshift peak is completely removed. For the irregular galaxy template the high redshift peak is significantly lowered and is now at a similar level to the low redshift peak. This analysis demonstrates that G2 is unlikely to be a field galaxy at $z =$~4.50 because the implied luminosity is too large ($M_{b_j}=-$21.95; 4~$L^*$). Furthermore, if G2 were at $z =$~4.50 then we would expect it to also be gravitationally lensed by G1. However, there is no evidence of a counter lensed image in either the Keck adaptive optics or the {\it Hubble Space Telescope} imaging.

The best fitting template for G2 is an elliptical galaxy (instantaneous burst) at $z_{phot} =$~0.86$^{+0.08}_{-0.21}$ (68 per cent confidence limits) with a young 0.5~Gyr stellar population and low dust content ($A_V = 0$). However, some caution is expressed over the properties of the best fitting template since there are known to be strong degeneracies between stellar age and dust extinction. It is intriguing that the most probable photometric redshift corresponds almost exactly to the redshift of the lensing galaxy G1 ($z=$~0.867). However, this may just be a fortunate coincidence. For example, the derived photometric redshift for G1 is 0.46$^{+0.41}_{-0.26}$ (68 per cent confidence limits), which is only just consistent with the spectroscopic redshift. Therefore, further photometric measurements of G2 will need to be carried out to establish the full observed spectral energy distribution, and produce a more robust estimate of the photometric redshift.

\section{Gravitational Lens Model}
\label{model}

In this section, we test whether G2 could be responsible for the flux-ratio anomaly observed between the three merging cusp images by constructing a new mass model for B2045+265. We assume that G2 is a dwarf satellite, or its remnant, {\it at the redshift} of G1, which is consistent with the results from the spectral energy distribution fitting above.

As galaxy G1 is well-fitted by a de Vaucouleurs profile, we assume it to be an early-type elliptical galaxy. In \citet{treu04} and \citet{koopmans06} it was shown, based on combined gravitational lensing and stellar kinematic data, that an isothermal mass model (i.e.\ $\rho(r) \propto r^{-2}$) often provides an excellent description of the smooth density profiles of massive early-type galaxies out to a redshift of one, within their inner 0.3--5 effective radii. Even though the stellar velocity dispersion of 213$\pm$23~km~s$^{-1}$ found by \citet{hamana05} indicates a somewhat more shallow density profile than isothermal, the source redshift identified by \citet{fassnacht99a} is still disputed. A higher source redshift would lead to a lower lensing galaxy mass and consequently a steeper density profile in order to match the observed stellar velocity dispersion. B2045+265 also exhibits a possible overdensity of nearby galaxies (e.g. \citealt{fassnacht99a}) about $\sim$10 arcsec west of the lens system.

We model the mass distribution of the system with two singular isothermal ellipsoids (SIE), one for G1 and one for G2, plus an external shear. The constraints on the mass model are provided by the positions of the lensed images and the radio-loud AGN within G1 (0.1\,mas accuracy) obtained from the new VLBA observations (see Table~\ref{rad-tab}), supplemented by the average flux-ratios obtained from a MERLIN Key-Program monitoring program ($\sim$2~per cent accuracy; \citealt{koopmans03}). In order not to underconstrain the model, we fix the mass centroid of G1 and G2 to the brightness centroid measured from the Keck adaptive optics imaging (see Table \ref{opt-tab}).

\begin{table}
\begin{center}
\caption{Lens model parameters for B2045+265. Given are the lens strength ($b$), position angle ($\theta$) and axial ratio ($q$), as well as the external shear strength and its position angle ($\gamma,\theta)_{\rm ext}$. The source position and normalized flux are given in the last three rows (the source flux $S_{\rm s}$ is normalized, assuming $F_{\rm A}\equiv 1$.). The second column indicates the most likely value, whereas the third column indicates the (marginalized) median value from the MCMC simulation and the 68 percent confidence levels.}
\begin{tabular}{lrrl} \hline

$b_{\rm G1}$	&        1.060		& 1.060$\pm$0.002 	& arcsec\\
$\theta_{\rm G1}$&	 22.3		& 22.4$\pm$0.1~~~~	& degree\\
$q_{\rm G1}$	&        0.66		& 0.66$\pm$0.01~~	& \\
\hline
$b_{\rm G2}$	&        0.087		& 0.087$\pm$0.003	& arcsec \\
$\theta_{\rm G2}$ &	 120.87		& 120.87$\pm$0.03~~	& degree\\
$q_{\rm G2}$	&        0.133		& 0.133$\pm$0.003	& \\
\hline
$\gamma_{\rm ext}$ &     0.215   	& 0.215$\pm$0.002	& \\
$\theta_{\rm ext}$ &     110.05		& 110.06$\pm$0.09~~	& degree\\
\hline
$x_{\rm s}$ &            $-$0.641	& $-$0.641$\pm$0.001 	& arcsec\\
$y_{\rm s}$ &            $-$0.654	& $-$0.654$\pm$0.001	& arcsec\\
$F_{\rm s}$ &            0.104		& 0.104$\pm$0.003	& \\

\hline
\end{tabular}
\label{mod-tab}
\end{center}
\end{table}

The SIE mass models of G1 and G2 are free to change their axial ratios, position angles and lens strengths. The shear strength and position angle are also free to change. The total number of free parameters is 11, whereas the number of constraints is 12. The number of degrees of freedom is therefore NDF~$=$~1 and the models should result in $\chi^2\simeq$~1.  The model parameters are quickly found by optimizing the $\chi^2$ penalty function, using the code described in {\citet{koopmans03b}, resulting in a best-fit model with $\chi^2/{\rm NDF}=$~1.9.\footnote{The density slope found by \citet{hamana05}, $\beta=$~1.58, gives $\chi^2=$~5.3. Leaving the density slope as a free parameter results in $\beta=$~1.8 with $\chi^2=$~0.0 (this case has NDF~$=$~0).} The resulting model of the system is shown in Fig.~\ref{best-model} and the best fitting parameters are listed in Table~\ref{mod-tab} with non-linear 68 per cent confidence levels obtained through a Markov-Chain Monte-Carlo simulation of 10$^6$ samples. Note that the confidence limits quoted in Table~\ref{mod-tab} are the formal uncertainties and the systematics associated with the choice of model are much larger.

We note several points. First, the model requires a significant external shear of $\sim$0.2. Although large, it is not unexpected given the overdensity of nearby galaxies in the western direction of the shear position angle. However, the position angle of G1 is offset by $\sim$90 degrees from that of the external shear, which could point to a compensating effect, i.e. the lens potential is more boxy than allowed by the elliptical mass model. Hence, part of the external shear might be the result of a non-elliptical halo around G1. Given the very tight constraints on the image positions and flux-ratios, these results are very robust (under the assumption that the mass model is chosen correctly). Second, the mass axial ratio of G2 is very small ($\sim$0.13). Such values can only be expected from either edge-on disk systems or possibly a tidally disrupted dwarf galaxy (see e.g. \citealt{li06}). Whether this is the correct interpretation of the model results remains to be tested with additional, deeper, observations. We note that although the mass axial ratio is more elongated than the light axial ratio measured from the F160W imaging presented in Section \ref{hst}, the position angle of the G2 elongation is consistent with the position angle of the light distribution (observed $q_{G2}=$~0.56$\pm$0.18 and $\theta_{G2}=$~106.3$\degr$$\pm$21.1$\degr$). Third, the resulting critical curve (Fig.~\ref{best-model}) is similar to that produced by {\citeauthor{congdon05} (2005; their Figure 4) near images A--C. Even though they use a completely different method to model the system, the protrusion of the critical curve near the cusp images is similar to what a dwarf disk galaxy, or a tidally disrupted dwarf galaxy (e.g. \citealt{li06}), would cause when situated near G2.

\begin{figure}
\begin{center}
\setlength{\unitlength}{1cm}
\begin{picture}(6,7.6)
\put(-2.45,-0.90){\includegraphics{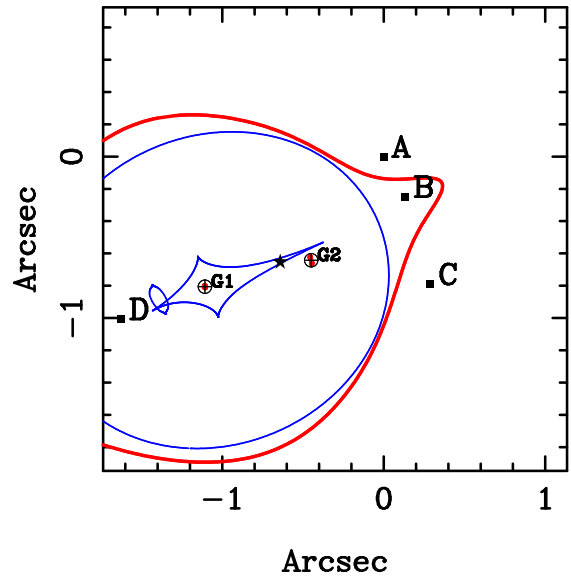}}
\end{picture}
\caption{The B2045+265 mass model which reproduces the relative positions and flux density ratios for the lensed images (A, B, C and D) without the need for additional cold dark matter substructure. The mass model consists of two singular isothermal ellipsoids representing G1 and G2, with an external shear. The background source position is marked by the star. The lens critical curve is shown by the thick (red) line. The thin (blue) line shows the source plane caustics. The model parameters are given in Table~\ref{mod-tab}.}
\label{best-model}
\end{center}
\end{figure}

\section{Discussion \& Conclusions}
\label{conclusions}

We have presented high-resolution radio, optical and infrared images of the gravitational-lens system CLASS B2045+265, obtained with the VLBA, {\it Hubble Space Telescope} and Keck (adaptive-optics) Telescope, respectively. In particular, the data were analyzed to examine the nature of the strong flux-ratio anomaly (e.g. \citealt{mao98}) between the three cusp images. This anomaly has been ascribed to the presence of CDM substructure around the main lensing galaxy (e.g. \citealt{dalal02}), as predicted from numerical simulations \citep{moore98,klypin99}. As such, these anomalies could be an important probe of the CDM paradigm.

We have shown the following. First, the lensed radio images, because of their compactness in the VLBA images, are not strongly affected by differential scattering at 5 GHz (e.g. \citealt{biggs04}), even though the images are scintillating due to the Galactic ionized medium \citep{koopmans03}. In addition, the flux-ratio anomaly observed at radio wavelengths (around 5~GHz) persists over at least eight months \citep{koopmans03}. The flux-ratio anomaly is therefore most likely due to a genuine perturbation of the smooth mass model of the lensing galaxy. Second, similar flux-ratios between the radio, optical and infrared further support that a mass perturbation in the lensing galaxy is causing the anomalies in this system, each of which individually could be affected by different processes (e.g.\ dust and microlensing in the optical, scattering in the radio, etc.) but unlikely in a way that is similar in magnitude over this wide range of wavelengths.

This leaves open the question what the true nature of the mass perturbation is. Are they caused by purely dark CDM satellites (e.g. \citealt{dalal02,kochanek04}), or by luminous CDM substructure in the form of normal dwarf satellites as we see them around the Milky Way and other galaxies?  

In the present Universe, luminous dwarf satellites might not be sufficient to explain these anomalies (e.g. \citealt{dalal02}). However, we expect their numbers to be more abundant at higher (lens) redshifts, simply because many of them will have merged with their hosts between the lens redshifts and $z\sim$~0. For example, the anomaly in MG~2016+112 \citep{koopmans02} is most likely caused by a nearby dwarf satellite detected through its [O~{\sc ii}] emission at the lens redshift (\citealt{koopmans02b}; see also \citealt*{kochanek04b} and \citealt{chen07}). Another example is MG~0414+053 where the addition of a luminous dwarf galaxy found close to the main lensing galaxy \citep{schechter93} results in a model which reproduces the observed positions and flux-densities of the lensed images (\citealt{trotter00}; \citealt{ros00}). However, the case of B2045+265 is one where the flux-ratio anomaly is very strong (e.g. \citealt{keeton03,metcalf05}) and therefore of particular interest when investigating the nature of these anomalies.

As in the cases of MG~2016+112 and MG~0414+053, we find evidence for the presence of at least one dwarf satellite G2 near the main lensing galaxy G1 (Figs.~\ref{ao} and \ref{hst-image}) in the {\it Hubble Space Telescope} images with WFPC2-F814W and NICMOS-F160W, and in the {\it K}-band Keck adaptive optics images. G2 resides in between the lensing galaxy and the cusp images. In addition, the lensing galaxy shows significant structure in the F814W-F160W colour images, suggesting the presence of starforming dwarf satellites. The [O~{\sc ii}] emission detected from this system supports this interpretation, but is not conclusive because it cannot be spatially matched (as in MG~2016+112). If the half-dozen distinct peaks in the F814W image around G1 are all dwarf satellites and not image artifacts, then this number is somewhat larger than the expected number of about three observable dwarf satellites, given that G2 is $\sim$1.5 mag above the 3--$\sigma$ limit for detection and the local luminosity function of dwarf satellites is $\propto L^{-1.8}$ (\citealt*{smith04}; we normalise the integral over the luminosity function above the luminosity of G2 to unity). Having built the case that G2 can be a luminous dwarf satellite of the lensing galaxy G1, we have presented a new lens model for B2045+265, taking this second brightest group member into account (G2). The result indicates that the satellite mass distribution must be significantly flattened, which could either be due to an edge-on disk or possibly a tidally disrupted dwarf galaxy being accreted on to G1 (see e.g. \citealt{li06} for the effect of these streams on lens image flux-ratios). We realize that this model is speculative at present. However, the lens-model critical curves are remarkably similar to an independent multi-pole expansion model of this system by \citet{congdon05}. 

It now seems likely that three out of the seven gravitational lens systems used to statistically detect dark matter substructure around lens galaxies \citep{dalal02} show evidence for CDM substructure in the form of luminous dwarf satellites. Furthermore, this luminous substructure appears to be more massive than expected when compared with the results from numerical simulations. For example, in the cases of B2045+265 and MG~2016+112 the luminous dwarf satellite constitutes about 1~per cent of the total halo mass, and for MG 0414+053 about 0.3~per cent. This mass-fraction, which is sensitive to the region contained within the lensed image positions, seems somewhat larger than to what is predicted for the inner few kpc of galaxy-scale haloes from numerical simulations i.e. $\leq$~0.5 per cent (\citealt{mao04}; \citealt*{diemand07}). This discrepancy between the observed and simulated mass-fraction could be due to an over-merging problem at the cores of simulated haloes or an over-estimation of the effect tidal stripping by baryons has on dark matter substructures in the numerical simulations. Alternatively, there could be additional unseen CDM substructure contributing to the flux-ratio anomaly which is not associated with the main lensing halo, but lies along the line of sight to the lensed images of the background quasar \citep{metcalf05b}.

In summary, the early-type lens galaxy in the system B2045+265 shows evidence for additional mass structure in the form of luminous dwarf satellites. The system also shows one of the most anomalous flux-ratios between its three cusp images of any lens system. Although we have not definitely shown that the dwarf satellite (G2) is responsible for the anomaly, the presence of significant structure in the brightness of G1 and the presence of G2, makes it more likely than previously assumed. A similar explanation was found for the anomalies in MG~2016+112 (\citealt{kochanek04b}; \citealt*{koopmans02b}; \citealt{chen07}) and MG~0414+053 \citep{trotter00,ros00}. Hence, deep high-resolution optical and infrared observations of each anomalous system is highly desirable to settle the question whether any or all anomalies are due to either luminous dwarfs or purely dark CDM substructures.

\section*{ACKNOWLEDGMENTS}
We thank Richard Porcas and Peter Schneider for useful comments and discussions. We also thank Hendrik Hildebrandt for his advice on calculating photometric redshifts. The Very Long Baseline Array is operated by the National Radio Astronomy Observatory which is a facility of the National Science Foundation operated under cooperative agreement by Associated Universities, Inc. The results presented herein were based on observations collected with the NASA/ESA {\it HST}, obtained at STScI, which is operated by AURA, under NASA contract NAS5-26555. These observations are associated with programme numbers 6629 and 9744. Some of the data presented herein were obtained at the W. M. Keck Observatory, which is operated as a scientific partnership among the California Institute of Technology, the University of California and the National Aeronautics and Space Administration. The Observatory was made possible by the generous financial support of the W. M. Keck Foundation. CEF acknowledges support from NSF's Research Experience for Undergraduates program, PHY-0243904. LVEK acknowledges partial support for this research from a VIDI grant (639.042.505) of the Netherlands Organisation for Scientific Research (NWO). DT and BTS were supported by the Spitzer Space Telescope project. DT is also supported by LTSA grant NRA-00-01-LTSA-064. This work is supported by the European Community's Sixth Framework Marie Curie Research Training Network Programme, Contract No. MRTN-CT-2004-505183 `ANGLES'.

\bsp

\label{lastpage}

\end{document}